\documentclass[twocolumn,floatfix,nofootinbib,superscriptaddress]{revtex4}
\usepackage{amssymb}
\usepackage{amsmath}
\usepackage{graphicx,subfigure,color,dcolumn,booktabs,bm}
\usepackage{longtable,lscape}
\usepackage{txfonts}
\usepackage{overpic}
\usepackage{indentfirst}
\usepackage{cases}
\usepackage{multirow}
\usepackage{ulem}
\usepackage[colorlinks,
            citecolor=blue,
            anchorcolor=red,
            menucolor=red,
            linkcolor=red,
            filecolor=red,
            runcolor=red,
            urlcolor=blue,
            frenchlinks=false]{hyperref}

\usepackage{soul}
\soulregister\cite7
\soulregister\ref7
\soulregister\eqref7
\soulregister\pageref7
\sethlcolor{yellow}

\newsavebox{\hlrevbox}

\allowdisplaybreaks
\begin{document}

\title{Mass relations in heavy hadrons from Jensen-like inequalities}

\author{Wen-Xuan Zhang}
\email{zhangwx89@outlook.com}
\affiliation{School of Physical Science and Technology, Lanzhou University, Lanzhou 730000, China}
\affiliation{Lanzhou Center for Theoretical Physics, Key Laboratory of Theoretical Physics of Gansu Province, Key Laboratory of Quantum Theory and Applications of MoE, Gansu Provincial Research Center for Basic Disciplines of Quantum Physics, Lanzhou University, Lanzhou 730000, China}
\affiliation{Research Center for Hadron and CSR Physics, Lanzhou University and Institute of Modern Physics of CAS, Lanzhou 730000, China}

\author{Wen-Nian Liu}
\affiliation{Institute of Theoretical Physics, College of Physics and Electronic Engineering, Northwest Normal University, Lanzhou 730070, China}
\affiliation{Xinjiang Laboratory Phase Transitions and Microstructures in Condensed Matters, College of Physical Science and Technology, Yili Normal University, Yining 835000, China}

\author{Duojie Jia}
\affiliation{Lanzhou Center for Theoretical Physics, Key Laboratory of Theoretical Physics of Gansu Province, Key Laboratory of Quantum Theory and Applications of MoE, Gansu Provincial Research Center for Basic Disciplines of Quantum Physics, Lanzhou University, Lanzhou 730000, China}
\affiliation{General Education Center, Qinghai Institute of Technology, Xining 810000, China}

\date{\today}

\begin{abstract}
We demonstrate that mass inequalities for hadrons with one or more heavy quarks arise primarily from the concavity of binding energies in the quark model, reflecting short-range Coulombic interactions and long-range confinement. Empirical two-body bindings $B_{i\bar{j}}$ are extracted from spin-averaged meson masses, ensuring model independence and direct use of experimental data. Fitting these as functions of reduced mass $\mu_{ij}$ reveals a critical confinement scale of 1.34~fm where bindings turn positive. The concave $B(1/\mu)$ motivates Jensen-like relations under flavor permutation, reproducing relations like $m_{x\bar{y}} > \frac{1}{2}(m_{x\bar{x}} + m_{y\bar{y}})$ and baryon analogs, including $m_{xyz} > \frac{1}{3}(m_{xxx} + m_{yyy} + m_{zzz})$. Hadron mass decomposition validates $\Delta M_{\textrm{EXP}} \approx \Delta B + \Delta C$ with $\sigma \sim 2.07$~MeV for mesons and baryons. Using the decomposition we obtain approximate mass estimates for unobserved heavy baryons (e.g., $M(\Omega_{b}^{\ast})\approx 6076.6\,$MeV, $M(\Xi_{cc}^{\ast})\approx 3703.6\,$MeV and $M(\Omega_{cc}^{\ast})\approx 3802.4\,$MeV) and identify favored quark-exchange scattering channels. The present analysis is phenomenological and does not constitute a rigorous three-body derivation.
\end{abstract}

\maketitle

\section{Introduction}
\label{sec:intro}

Understanding the nonperturbative regime of strong interactions remains a central challenge in quantum chromodynamics (QCD). At short distances, the interaction exhibits Coulomb-like behavior with asymptotic freedom, potentially leading to concave forms in binding energies from inverse-mass dependence. In contrast, at long ranges, confinement manifests as a linear potential, driving string-like bindings that may break at critical scales, introducing strong condensation and unquenched effects and complicating mass predictions for heavy hadronic states. This is timely given the recent LHCb discovery of the doubly-charmed baryon $\Xi_{cc}^+$~\cite{LHCb:2026pxn} and the surge in triply-heavy baryon mass predictions~\cite{Zhou:2025fpp}.

Empirical mass inequalities among hadrons containing heavy or strange quarks provide valuable insights into these dynamics. Mixed-flavor mesons and baryons systematically exceed the mean value of same-flavor counterparts, as in
\begin{equation}
    m_{x\bar{y}} > \frac{1}{2}(m_{x\bar{x}}+m_{y\bar{y}}), \quad
    m_{xxy} > \frac{1}{2}(m_{xxx}+m_{xyy}),
    \label{equ:inequality}
\end{equation}
with $x \neq y$ and at least one heavy flavor~\cite{Bertlmann:1979zs,Richard:1983mu,Weingarten:1983uj,Witten:1983ut,Nussinov:1983hb,Lieb:1985aw,Martin:1986da,Anwar:2017toa,Karliner:2019vhw}. Here the symbols $x$ and $y$ denote quark flavor labels (e.g., $c$ for charm, $b$ for bottom), not numerical variables; the relations are applied to the mass function evaluated at these discrete labels. These relations, valid for pseudoscalar, vector mesons as well as ground-state baryons, indicate concavity in the mass function with respect to energy scale, stemming from flavor-dependent interactions in the quark model. Similar mass decompositions into constituent quark masses, two-body bindings, and hyperfine (chromomagnetic) terms have been widely used for heavy hadrons and tetraquarks \cite{Karliner:2014gca,Karliner:2017elp,Karliner:2017qjm,Ebert:2004ck,Anwar:2017toa,Yang:2019lsg,Liu:2019zoy,wu:2016vtq,luo:2017eub,chen:2022asf,Cheng:2020wxa,li:2023aui}.

In this work, we interpret these inequalities phenomenologically in terms of the concavity of chromoelectric binding energies $B_{i\bar{j}}$, extracted empirically from spin-averaged hadron masses to ensure model independence and reliance on experimental data. By fitting $B(\mu)$ and $B(1/\mu)$, interpreted as the mean values of the Cornell potential, we explore their concave dependence on reduced mass $\mu_{ij}$, yielding Jensen-like relations under flavor permutation. This fitting further reveals a critical confinement scale upon positive bindings, aligning with predictions for string-breaking distances. We validate the approach with $\Delta M_{\textrm{EXP}} \approx \Delta B + \Delta C$, and obtain approximate mass estimates for unobserved doubly and triply heavy baryons, and discuss potential decay and scattering processes. This phenomenological study focuses on the empirical relations of interaction energies extracted from experimental meson masses, rather than pursuing a fully rigorous three-body derivation.

\section{Mass inequalities and their origin}
\label{sec:origin}

The mass inequalities (\ref{equ:inequality}) can be interpreted using a Jensen-like relation for a concave function $f$,
\begin{equation}
    f(\sum_{i}^{n}\alpha_{i}x_{i}) > \sum_{i}^{n}\alpha_{i}f(x_{i}),
    \label{equ:Jensen-like}
\end{equation}
with weights $\alpha_{i}$ summing to 1. For $n=2$, $\alpha_{i} = 1/2$, and flavor labels $x_1 = x$, $x_2 = y$ one obtains the meson and baryon relations in Eq.~(\ref{equ:inequality}). For $n=3$, weights $\alpha_{i} = 1/3$ yield $m_{xyz} > \frac{1}{3}(m_{xxx} + m_{yyy} + m_{zzz})$. These relations hold for mesons with at least one heavy quark and baryons where the $xy$ pair exceeds the $ud$ diquark mass \cite{Karliner:2019vhw}. Such patterns arise from concavity of flavor-dependent interactions \cite{Nussinov:1983hb,Lieb:1985aw,Martin:1986da}, particularly when exchanging heavy-light flavors, indicating flavor symmetry breaking.

The concavity of the mass function can be motivated from the underlying quark model dynamics. To illustrate this, consider the hadron mass in second-order perturbation theory:
\begin{align}
    H = H_0 + \lambda V&, \quad H_0 |\psi_0\rangle = E_0 |\psi_0\rangle, \nonumber \\
    \textrm{d}_{\lambda}E_0 = \langle\psi_0|V|\psi_0\rangle&, \quad
    \textrm{d}^2_{\lambda}E_0 = 2 \sum_{n\neq 0} \frac{|\langle\psi_n|V|\psi_0\rangle|^2}{E_0-E_n} \leq 0,
    \label{equ:soptheorem}
\end{align}
where $H_0$ is the leading-order Hamiltonian, $V$ denotes perturbative interactions, and the variational parameter $\lambda$ is associated with the inverse reduced mass $1/\mu$. The concavity can be examined by analyzing the second derivatives of the mass with respect to the parameter $\lambda$, providing a phenomenological basis for the observed Jensen-like pattern. In the quark model, the mass incorporates both leading-order and perturbative interactions:
\begin{equation}
    m = \sum_{i} m_{i} + E_{\textrm{BD}} + E_{\textrm{CMI}},
    \label{equ:mass}
\end{equation}
where
\begin{align}
    &E_{\textrm{BD}} = \sum_{i<j} B_{ij(\bar{j})}, 
    \label{equ:massBD} \\
    &E_{\textrm{CMI}}=-\sum_{i<j} \langle\mathbf{\lambda}_{i}\cdot\mathbf{\lambda}_{j}\rangle 
    \langle\mathbf{\sigma}_{i}\cdot\mathbf{\sigma}_{j}\rangle C_{ij(\bar{j})}. 
    \label{equ:massCMI}
\end{align}
Here, $E_{\textrm{BD}}$ captures confinement and short-range chromoelectric binding, while $E_{\text{CMI}}$ represents the perturbative chromomagnetic interaction. The binding energies $B_{ij(\bar{j})}$ are regarded as mean values of the Cornell potential for $\mathbf{1}_c$ ($i\bar{j}$) and $\mathbf{\bar{3}}_c$ ($ij$). The meson coupling parameters $C_{i\bar{j}}=A_{i\bar{j}}/m_{i}m_{j}$ are taken from Ref.~\cite{Zhang:2025wmr}, with $A_{i\bar{j}}$ fitted to a convex function of the reduced mass $\mu$, while the baryon couplings are set as $C_{ij}=0.85\,C_{i\bar{j}}$ (see Sec.~\ref{sec:binding}). The constituent masses used for $\mu$ are extracted from the Regge-like framework \cite{Zhang:2025wmr,Jia:2019bkr}. The bindings $B_{ij(\bar{j})}$ and couplings $C_{i\bar{j}}$ themselves are fixed from experimental meson masses and hyperfine splittings and do not depend on this mass choice. Further investigation of concavity of interaction energies follows from the condition (\ref{equ:soptheorem}) when $\lambda$ is associated with $1/\mu$, leading to the fitting of binding energies $B(1/\mu)$.

From another perspective, the mass inequalities (\ref{equ:inequality}) can be related to Eqs.(\ref{equ:mass})--(\ref{equ:massCMI}), by decomposing the differences:
\begin{align}
    &\Delta M = m(x\bar{y})-\frac{1}{2}[m(x\bar{x})+m(y\bar{y})] \nonumber \\ 
    &\Delta B = E_{\textrm{BD}}(x\bar{y})-\frac{1}{2}[E_{\textrm{BD}}(x\bar{x})+E_{\textrm{BD}}(y\bar{y})] \nonumber \\
    &\Delta C = E_{\textrm{CMI}}(x\bar{y})-\frac{1}{2}[E_{\textrm{CMI}}(x\bar{x})+E_{\textrm{CMI}}(y\bar{y})] \nonumber \\
    &\Delta M = \Delta B + \Delta C,
    \label{equ:masstransform}
\end{align}
This decomposition enables us to examine the mass inequalities and therefore obtain approximate mass relations by evaluating $\Delta B$ and $\Delta C$.

\section{Concavity of binding energies}
\label{sec:binding}

The binding energy is primarily contributed by short-range chromoelectric interactions within heavy quarks or heavy-strange systems, as considered in various works \cite{Karliner:2014gca,Karliner:2017elp,Karliner:2017qjm,Karliner:2020vsi,Song:2022csw,Liu:2023vrk,Zhang:2025wmr}. Ref.\cite{Karliner:2014gca} employed five bindings $B_{cs}$, $B_{bs}$, $B_{cc}$, $B_{bc}$ and $B_{bb}$ to study doubly heavy systems. In the QCD string picture \cite{Song:2022csw,Liu:2023vrk,Zhang:2025wmr}, which incorporates perturbative gluon interactions and confinement, we extract bindings from the spin-averaged masses of mesons:
\begin{equation}
    B_{i\bar{j}} = \bar{M}_{i\bar{j}}-\bar{M}_{i\bar{n}}-\bar{M}_{n\bar{j}}+\bar{M}_{n\bar{n}}.
    \label{equ:extractBD}
\end{equation}
Here $\bar{M}$ denotes the spin-averaged mass of the meson, and the light non-strange flavor is labeled by $n$ ($u$ or $d$).
The mass of 686.9$\,$MeV for the pseudoscalar meson $\eta_s$ is solved via \cite{Witten:1983ut}
\begin{equation}
    m_{\eta_s} = (m_{K^+}^2+m_{K^0}^2-m_{\pi^\pm}^2)^{1/2}
    \label{equ:massetas}
\end{equation}
supported by lattice QCD simulations \cite{Davies:2009tsa,Blossier:2010cr,HPQCD:2011qwj,Davies:2012xun}. The mass of 6332$\,$MeV for the vector meson $B_c^{\ast}$ is adopted from \cite{Ebert:2002pp}, supported by lattice QCD simulations \cite{Gregory:2009hq,Mathur:2016hsm,Mathur:2018epb}. The values of binding energies are listed collectively in Table \ref{tab:bindings}, including a potential binding of $-40.6\,$MeV within $s\bar{s}$, where $s$ and $\bar{s}$ are both massive compared to non-strange light flavors. For baryon systems, the binding energies $B_{ij}$ under the $\mathbf{\bar{3}}_c$ representation can be obtained by scaling with color factors, $B_{ij} = B_{i\bar{j}}/2$, following the principle of chromoelectric interactions.

\renewcommand{\tabcolsep}{0.3cm}
\renewcommand{\arraystretch}{1.2}
\begin{table}[!htbp]
    \caption{The binding energies $B_{i\bar{j}}$ and $B_{ij}$ and coupling parameters $C_{i\bar{j}}$ and $C_{ij}$ with their associated reduced masses $\mu_{ij}$. All parameters are in unit of MeV.}
    \label{tab:bindings}
    \begin{tabular}{crrrrr}
        \toprule[1.5pt]\toprule[0.5pt]
        $ij$ & $\mu_{ij}$ & $B_{i\bar{j}}$ & $B_{ij}$ & $C_{i\bar{j}}$ & $C_{ij}$ \\ \hline
        $nn$ & 115.0 & 0.0 & 0.0 & 30.0 & 25.5 \\
        $sn$ & 135.2 & 0.0 & 0.0 & 18.7 & 15.9 \\
        $ss$ & 164.0 & $-40.6$ & $-20.3$ & 15.6 & 13.2 \\
        $cn$ & 198.3 & 0.0 & 0.0 & 6.7 & 5.7 \\
        $bn$ & 218.8 & 0.0 & 0.0 & 2.1 & 1.8 \\
        $cs$ & 267.2 & $-76.0$ & $-38.0$ & 6.7 & 5.7 \\
        $bs$ & 305.6 & $-91.0$ & $-45.5$ & 2.3 & 1.9 \\
        $cc$ & 720.0 & $-258.9$ & $-129.4$ & 5.3 & 4.5 \\
        $bc$ & 1089.7 & $-352.0$ & $-176.0$ & 2.7 & 2.3 \\
        $bb$ & 2240.0 & $-566.8$ & $-283.4$ & 2.9 & 2.5 \\
        \bottomrule[0.5pt]\bottomrule[1.5pt]
    \end{tabular}
\end{table}

\begin{figure}[t]
    \centering
    \includegraphics[width=0.48\textwidth]{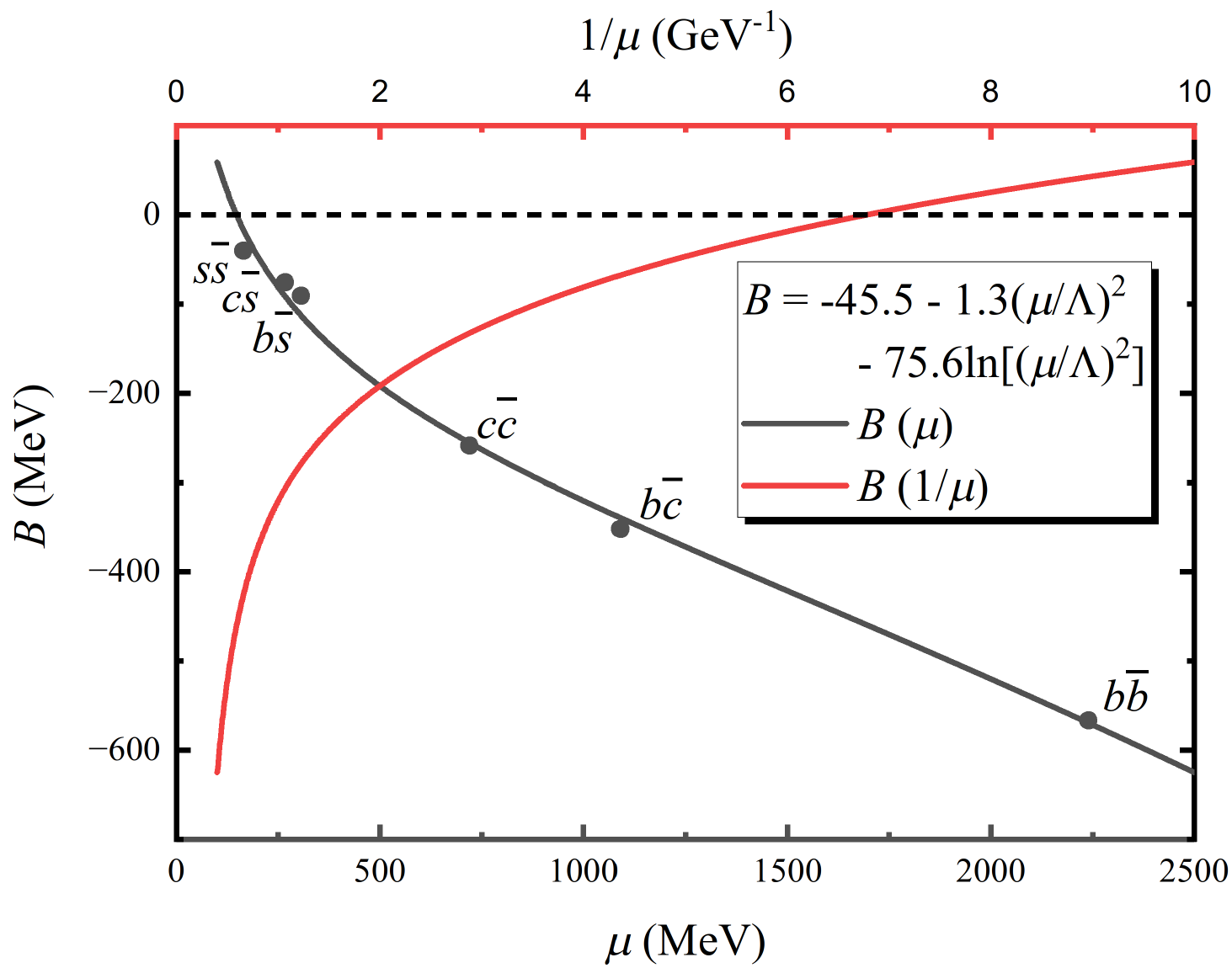}
    \caption{Plot of fitted binding energies for meson systems with respect to the reduced mass $\mu$ of quark-antiquark pair. The gray curve denotes Eq.(\ref{equ:bindingfit}) with discrete data from Table \ref{tab:bindings} shown as spherical dots. The red curve represents Eq.(\ref{equ:bindingfitR}) as a function of $1/\mu$.}
    \label{fig:BqqbarFit}
\end{figure}

Given the effective (non-vanishing) binding energies in Table \ref{tab:bindings}, we obtain the fitted relations with respect to the reduced mass $\mu_{ij} = m_i m_j/(m_i + m_j)$ for meson systems: 
\begin{align}
    &B(\mu) = -45.5 - 1.3 \frac{\mu^2}{\Lambda^2} - 75.6 \textrm{ln}(\frac{\mu^2}{\Lambda^2}), 
    \label{equ:bindingfit} \\
    &B(1/\mu) = -45.5 - 1.3 \frac{1}{\Lambda^2/\mu^2} - 75.6 \textrm{ln}(\frac{1}{\Lambda^2/\mu^2}),
    \label{equ:bindingfitR}
\end{align}
where $\Lambda=200\,$MeV is the QCD scale, the masses $m_i$, $m_j$ of flavors are as given below Eq.~(\ref{equ:massCMI}), and the function bases are chosen as in Ref.\cite{Zhang:2025wmr}. A least-squares fit of this three-parameter form to the six effective binding energies yields an rms residual of $\approx 15.8\,$MeV and $\chi^{2}/\mathrm{dof}\approx 497$ with unit MeV weights. The large unweighted $\chi^{2}$ reflects the simplicity of the parametrization and the scatter in the light sector; the form is used to demonstrate the concave trend in $B(1/\mu)$, not as a precision mass formula. Applying the condition (\ref{equ:soptheorem}) with $\lambda$ associated with $1/\mu$, the concavity of binding energies can be examined in second derivatives, as well as in Fig.(\ref{fig:BqqbarFit}), serving as a phenomenological motivation for the observed mass patterns. Another contribution is from chromomagnetic interactions, where $A_{i\bar{j}}$ exhibits concavity with respect to $1/\mu$ and convexity with respect to $\mu$ in Ref.\cite{Zhang:2025wmr},

As shown in Fig.(\ref{fig:BqqbarFit}), binding energies should be dominated by Coulombic potential at very short distance, with contributions from the long-range confinement potential, indicating agreement with theoretical expectations and fitted data. At a critical reduced mass of 147.4$\,$MeV the binding energy becomes positive and loses its effectiveness. Converting with $\hbar c\approx 197.3\,$MeV\,fm gives a length $197.3/147.4\approx 1.34\,$fm, identified with the string-breaking distance. Lattice QCD and other studies predict this to be 1.2--1.4$\,$fm \cite{Bali:1998de,Bali:2005fu,Castorina:2007eb,Bulava:2019iut,Bonati:2020orj,Chagdaa:2021hul,Jiang:2023lmj,Kou:2024dml}, with the upper limit from Ref.\cite{Kou:2024dml} close to our result. This scale implies strong condensation and unquenched effects, leading to the instability of pseudo-Goldstone mesons such as $s\bar{n}$ and $n\bar{n}$.

Following the mass relations (\ref{equ:masstransform}), one can evaluate $\Delta B$ and $\Delta C$, to confirm that the mass patterns are consistent with binding energies and chromomagnetic interactions. The values of $C_{i\bar{j}}$ are extracted from meson data, and a ratio $C_{ij} = 0.85C_{i\bar{j}}$ is adopted to minimize the standard error $\sigma$. The traditional ratio $2/3$ used in early quark-model analyses \cite{Lipkin:1986dx,Keren-Zur:2007ytk,Liu:2019zoy} is a model hypothesis for meson/baryon hyperfine scaling rather than a requirement of the present decomposition; here $r=C_{ij}/C_{i\bar{j}}$ is fixed by fitting $\Delta M_{\textrm{EXP}}\approx\Delta B+\Delta C$. Scanning $r$ shows that $r=0.85$ yields the minimal $\sigma=2.07\,$MeV, while $r=2/3$ and $r=1$ give $\sigma\approx4.5$ and $4.4\,$MeV, respectively. Varying $r$ within $[0.75,0.95]$ shifts the masses by only a few MeV for doubly heavy states and by $\lesssim 15\,$MeV for $\Omega_{ccc}$---well within the typical theoretical errors among quark-model and lattice estimates. Comparisons are shown explicitly in Table~\ref{tab:proof}, where $\Delta B + \Delta C$ closely tracks $\Delta M_{\textrm{EXP}}$ for baryons, with slight errors for mesons due to isospin splittings. The small standard error $\sigma = 2.07\,$MeV confirms that the observed patterns are consistent with flavor-dependent interactions.

\renewcommand{\tabcolsep}{0.4cm}
\renewcommand{\arraystretch}{1.2}
\begin{table}[!htbp]
    \caption{Mass differences with experimental values of $\Delta M_{\textrm{EXP}}$ (in MeV) and the evaluated $\Delta B+\Delta C$ (in MeV). Mesons are used for fitting data to predict baryon properties. The standard error is determined to be $\sigma=2.07\,\text{MeV}$.}
    \label{tab:proof}
    \begin{tabular}{lcrr}
        \toprule[1.5pt]\toprule[0.5pt]
        Difference & $J^P$ & $\Delta M_{\textrm{EXP}}$ & $\Delta B+\Delta C$ \\ \hline
        $K^{\ast}<(\phi+\rho)/2$ & $1^-$ & $-3.8$ & $-1.8$ \\
        $D^{\ast}>(\mathrm{J}/\psi+\rho)/2$ & $1^-$ & 72.5 & 70.8 \\
        $D>(\eta_c+\pi)/2$ & $0^-$ & 307.8 & 305.4 \\
        $B^{\ast}>(\Upsilon+\rho)/2$ & $1^-$ & 206.9 & 207.0 \\
        $B>(\eta_b+\pi)/2$ & $0^-$ & 512.7 & 512.6 \\
        $D_s^{\ast}>(\mathrm{J}/\psi+\phi)/2$ & $1^-$ & 54.0 & 54.0 \\
        $B_s^{\ast}>(\Upsilon+\phi)/2$ & $1^-$ & 175.5 & 175.5 \\
        $B_c^{\ast}>(\Upsilon+\mathrm{J}/\psi)/2$ & $1^-$ & 53.4 & 53.4 \\
        $B_c>(\eta_b+\eta_c)/2$ & $0^-$ & 83.2 & 83.2 \\
        \midrule[0.5pt]
        $\Xi^{\ast}>(\Omega+\Sigma^{\ast})/2$ & $3/2^+$ & 4.7 & 0.8 \\
        $\Xi_c^{\ast}>(\Omega_c^{\ast}+\Sigma_c^{\ast})/2$ & $3/2^+$ & 3.5 & 0.8 \\
        $\Xi_c^{\prime}>(\Omega_c+\Sigma_c^{\prime})/2$ & $1/2^+$ & 3.9 & 0.8 \\
        $\Xi_c^{\prime}>(\Xi+\Xi_{cc})/2$ & $1/2^+$ & 110.0 & 109.5 \\
        $\Xi_b^{\prime}>(\Omega_b+\Sigma_b^{\prime})/2$ & $1/2^+$ & 4.6 & 0.8 \\
        \bottomrule[0.5pt]\bottomrule[1.5pt]
    \end{tabular}
\end{table}

The $K^{\ast}$ entry in Table~\ref{tab:proof} is the only case with $\Delta M_{\textrm{EXP}}<0$. Light mesons are governed mainly by long-range confinement, where the potential grows approximately linearly and no clear concavity emerges. The Jensen-like inequality is therefore not expected to hold in the same way as for systems with at least one heavy quark, which probe short-distance Coulombic binding. Both the experimental value ($-3.8\,$MeV) and the decomposition ($-1.8\,$MeV) are expected close to zero, due to the lack of binding contrast. The small negative sign is thus naturally attributed to experimental and theoretical uncertainties, rather than to a robust inverted inequality. All other entries involve heavy flavors, for which the concave binding pattern dominates and the inequalities are satisfied. The $K^{\ast}$ row is retained as a light-sector boundary check and does not affect the heavy-hadron predictions.

The four baryon rows in Table~\ref{tab:proof} that exchange $s$ and $n$ ($\Xi^{\ast}$, $\Xi_c^{\ast}$, $\Xi_c^{\prime}$, $\Xi_b^{\prime}$) all give $\Delta B+\Delta C=0.8\,$MeV, while experiment yields $3.5$--$4.7\,$MeV. With $B_{nn}=B_{sn}=0$, the $s$--$n$ exchange carries no binding contrast in those channels, so the few-MeV underestimation is a light-sector effect. It can arise from (i) the schematic vanishing of the bindings, (ii) a small residual in $s$--$n$ systems from the strange-quark mass and infrared confinement not captured when $B_{sn}=0$, and (iii) hyperfine uncertainties in $C_{ij}$. It does not indicate a failure of the heavy-quark concavity mechanism. The same $s$--$n$ pattern appears in $\Xi_b^{\ast}>(\Omega_b^{\ast}+\Sigma_b^{\ast})/2$ used for $M(\Omega_b^{\ast})$.

\section{Approximate mass relations}
\label{sec:prediction}

Using the binding energies $B_{ij(\bar{j})}$ and coupling parameters $C_{ij(\bar{j})}$ given in Table \ref{tab:bindings}, we evaluate the decomposition $\Delta B + \Delta C \approx \Delta M$ to obtain approximate numerical mass relations. This phenomenological approach allows us to estimate the masses for unobserved heavy baryons, as well as potential decay and scattering channels. Table~\ref{tab:equality} lists these estimates for doubly and triply heavy baryons, including extensions to relations involving averages over three baryons. These relations hold for the same spin configurations on both sides to avoid the interference from spin multiplets.

\renewcommand{\tabcolsep}{0.5cm}
\renewcommand{\arraystretch}{1.2}
\begin{table}[!htbp]
    \caption{Approximate mass differences suggested for unobserved baryons such as doubly and triply heavy baryons. The relation $\Delta M \approx \Delta B + \Delta C$ is used to estimate the differences (in MeV).}
    \label{tab:equality}
    \begin{tabular}{lcr}
        \toprule[1.5pt]\toprule[0.5pt]
        Relation & $J^P$ & $\Delta B + \Delta C$ \\ \hline
        $\Xi_c^{\ast}>(\Xi^{\ast}+\Xi_{cc}^{\ast})/2$ & $3/2^+$ & 28.5 \\
        $\Xi_b^{\ast}>(\Xi^{\ast}+\Xi_{bb}^{\ast})/2$ & $3/2^+$ & 90.5 \\
        $\Xi_b^{\ast}>(\Omega_b^{\ast}+\Sigma_{b}^{\ast})/2$ & $3/2^+$ & 0.8 \\
        $\Xi_b^{\prime}>(\Xi+\Xi_{bb})/2$ & $1/2^+$ & 201.9 \\
        $\Omega_c^{\ast}>(\Omega+\Omega_{cc}^{\ast})/2$ & $3/2^+$ & 28.5 \\
        $\Omega_b^{\ast}>(\Omega+\Omega_{bb}^{\ast})/2$ & $3/2^+$ & 90.5 \\
        \midrule[0.5pt]
        $\Xi_{cc}^{\ast}>(\Omega_{ccc}+\Sigma_{c}^{\ast})/2$ & $3/2^+$ & 39.8 \\
        $\Omega_{cc}^{\ast}>(\Omega_{ccc}+\Omega_{c}^{\ast})/2$ & $3/2^+$ & 28.5 \\
        $\Omega_{ccb}^{\ast}>(\Omega_{ccc}+\Omega_{cbb}^{\ast})/2$ & $3/2^+$ & 27.3 \\
        $\Xi_{bb}^{\ast}>(\Omega_{bbb}+\Sigma_{b}^{\ast})/2$ & $3/2^+$ & 109.2 \\
        $\Omega_{bb}^{\ast}>(\Omega_{bbb}+\Omega_{b}^{\ast})/2$ & $3/2^+$ & 90.5 \\
        $\Omega_{bbc}^{\ast}>(\Omega_{bbb}+\Omega_{bcc}^{\ast})/2$ & $3/2^+$ & 27.3 \\
        $\Xi_{bc}^{\ast}>(\Omega_{bcc}^{\ast}+\Sigma_{b}^{\ast})/2$ & $3/2^+$ & 39.8 \\
        $\Xi_{bc}>(\Omega_{bcc}+\Sigma_{b})/2$ & $1/2^+$ & 12.9 \\
        $\Omega_{bc}^{\ast}>(\Omega_{bcc}^{\ast}+\Omega_{b}^{\ast})/2$ & $3/2^+$ & 28.5 \\
        $\Omega_{bc}>(\Omega_{bcc}+\Omega_{b})/2$ & $1/2^+$ & 1.0 \\
        \midrule[0.5pt]
        $\Xi_{c}^{\ast}>(\Omega_{ccc}+\Omega+\Delta)/3$ & $3/2^+$ & 69.0 \\
        $\Xi_{b}^{\ast}>(\Omega_{bbb}+\Omega+\Delta)/3$ & $3/2^+$ & 200.5 \\
        $\Xi_{bc}^{\ast}>(\Omega_{bbb}+\Omega_{ccc}+\Delta)/3$ & $3/2^+$ & 176.3 \\
        $\Omega_{bc}^{\ast}>(\Omega_{bbb}+\Omega_{ccc}+\Omega)/3$ & $3/2^+$ & 146.3 \\
        \bottomrule[0.5pt]\bottomrule[1.5pt]
    \end{tabular}
\end{table}

The values of $\Delta B+\Delta C$ in Table~\ref{tab:equality} span a wide range, from $0.8$ to $201.9\,$MeV. This spread is expected from the concavity of $B(\mu)$. Large differences arise when a heavy quark is exchanged with a much lighter flavor (e.g., $\Xi_b^{\prime}>(\Xi+\Xi_{bb})/2$ or the $n=3$ routes involving $\Omega_{bbb}$), so that the concave binding is sampled over a wide interval in $1/\mu$ and the binding contrast dominates. On the contrary, the smallest entries---such as $\Xi_b^{\ast}>(\Omega_b^{\ast}+\Sigma_b^{\ast})/2$ giving $0.8\,$MeV and $\Omega_{bc}>(\Omega_{bcc}+\Omega_b)/2$ giving $1.0\,$MeV---correspond to $s$--$n$ exchanges. With $B_{nn}=B_{sn}=0$, those channels carry essentially no binding contrast, so $\Delta B+\Delta C$ is expected to lie close to zero. The residual few MeV arise mainly from hyperfine pieces and the light-sector systematics already discussed after Table~\ref{tab:proof}. The $\sim 0.8\,$MeV cases remain physically meaningful, alligning with the few-MeV $s$--$n$ underestimation noted above.

\renewcommand{\tabcolsep}{0.1cm}
\renewcommand{\arraystretch}{1.2}
\begin{table}[!htbp]
    \caption{Comparisons of approximate mass estimates (in MeV) obtained from the decomposition with other results from quark model \cite{Karliner:2014gca}, relativistical quark model \cite{Ebert:2004ck}, constituent quark model \cite{Yang:2019lsg}, and lattice QCD \cite{Brown:2014ena}. $M(\Omega_{b}^{\ast})$ was also predicted to be 6088$\,$MeV \cite{Ebert:2005xj}.}
    \label{tab:predictions}
    \begin{tabular}{crrrrr}
        \toprule[1.5pt]\toprule[0.5pt]
        Baryon & This work & \cite{Karliner:2014gca} & \cite{Ebert:2004ck} & \cite{Yang:2019lsg} & \cite{Brown:2014ena} \\ \hline
        $\Omega_{b}^{\ast}$ & 6076.6 & 6066.7 & \dots & \dots & 6085(47)(20) \\
        $\Xi_{cc}^{\ast}$ & 3703.6 & 3690(12) & 3727 & \dots & 3692(28)(21) \\
        $\Xi_{bb}^{\ast}$ & 10197.8 & 10184(12) & 10202 & \dots & 10178(30)(24) \\
        $\Xi_{bb}$ & 10151.3 & 10162(12) & 10237 & \dots & 10143(30)(23)  \\
        $\Omega_{cc}^{\ast}$ & 3802.4 & \dots & 3872 & \dots & 3822(20)(22) \\
        $\Omega_{ccc}$ & 4827.0 & \dots & \dots & 4798 & 4796(8)(18) \\
        $\Omega_{bbb}$ & 14360.0 & \dots & \dots & 14396 & 14366(9)(20) \\
        \bottomrule[0.5pt]\bottomrule[1.5pt]
    \end{tabular}
\end{table}

For the relation $\Xi_c^{\prime}>(\Xi+\Xi_{cc})/2$ in Table \ref{tab:proof}, one can explicitly estimate the mass of $\Xi_{cc}$ to be 3623.6$\,$MeV, consistent with the reported value 3621.6$\,$MeV \cite{LHCb:2017iph,LHCb:2018pcs}. Similarly, more estimates can be made if other hadrons are discovered experimentally, such as
\begin{align}
    &\Xi_b^{\ast}>(\Omega_b^{\ast}+\Sigma_{b}^{\ast})/2 \implies M(\Omega_{b}^{\ast})\approx 6076.6, \nonumber \\
    &\Xi_c^{\ast}>(\Xi^{\ast}+\Xi_{cc}^{\ast})/2 \implies M(\Xi_{cc}^{\ast})\approx 3703.6, \nonumber \\
    &\Xi_b^{\ast}>(\Xi^{\ast}+\Xi_{bb}^{\ast})/2 \implies M(\Xi_{bb}^{\ast})\approx 10197.8, \nonumber \\
    &\Xi_b^{\prime}>(\Xi+\Xi_{bb})/2 \implies M(\Xi_{bb})\approx 10151.3, \nonumber \\
    &\Omega_c^{\ast}>(\Omega+\Omega_{cc}^{\ast})/2 \implies M(\Omega_{cc}^{\ast})\approx 3802.4, \nonumber \\
    &\Xi_{c}^{\ast}>(\Omega_{ccc}+\Omega+\Delta)/3 \implies M(\Omega_{ccc})\approx 4827.0, \nonumber \\
    &\Xi_{b}^{\ast}>(\Omega_{bbb}+\Omega+\Delta)/3 \implies M(\Omega_{bbb})\approx 14360.0.
    \label{equ:predictions}
\end{align}
Comparisons are listed in Table \ref{tab:predictions} with some notable references.
There are various relations to predict the mass of $M(\Omega_{ccc})$, but only the Eq.~(\ref{equ:predictions}) is regarded for now. Chained $n=2$ routes for $M(\Omega_{ccc})$ via $M(\Omega_{cc}^{\ast})$ or $M(\Xi_{cc}^{\ast})$ differ from the direct $n=3$ value by up to $\sim 45\,$MeV, since residual errors are enlarged by the $1/2$ and $1/3$ inversion factors and accumulate under chaining. We therefore quote only the single-step $n=3$ estimates from measured $\Xi_{c}^{\ast}$ and $\Xi_{b}^{\ast}$ in Eq.~(\ref{equ:predictions}).

The other mass relations are expected to be confirmed in future experiments, both in mass spectra and scattering channels. For example, the inequality $D>(\eta_c+\pi)/2$ implies a scattering channel $D\bar{D} \to \eta_c + \pi$, with a mass gap of 307.8$\,$MeV which indicates a favored process. Baryon scatterings could be discovered in channels from Table \ref{tab:equality}, such as $\Xi_c^{\ast}\Xi_c^{\ast}\to\Xi^{\ast}+\Xi_{cc}^{\ast}$ (exchanging $c$--$n$), $\Xi_b^{\ast}\Xi_b^{\ast}\to\Xi^{\ast}+\Xi_{bb}^{\ast}$ (exchanging $b$--$n$), and $\Xi_{cc}^{\ast}\Xi_{cc}^{\ast}\to\Omega_{ccc}+\Sigma_{c}^{\ast}$ (exchanging $c$--$s$), i.e., by exchanging a heavy quark with a light or strange quark. This phenomenon aligns with the quark-exchange model for hadron scatterings \cite{Barnes:2000hu,Hilbert:2007hc,Yamaguchi:2019djj}, as the strong bindings (\ref{equ:massBD}) favor exchanging of heavy quarks or heavy-strange systems.

\section{Summary and conclusions}
\label{sec:summary}

In this work, we investigate a set of mass inequalities for hadrons containing one or more heavy quarks, interpreting them in terms of the concavity of binding energies $B_{ij(\bar{j})}$ as the mean values of the Cornell potential. By employing the quark model and decomposing the hadron mass into leading-order, binding contributions (\ref{equ:massBD}) from confinement, and chromomagnetic interactions (\ref{equ:massCMI}), we establish a mass relation $\Delta M \approx \Delta B + \Delta C$ under flavor permutations. This phenomenological approach yields Jensen-like relations for distinct flavors, such as $m_{xyz} > \frac{1}{3}(m_{xxx} + m_{yyy} + m_{zzz})$ for averages among three baryons, attributing the concavity to flavor-dependent interactions with respect to $1/\mu$ in second-order perturbation theory.

We extract effective two-body binding energies $B_{i\bar{j}}$ and hyperfine couplings $C_{i\bar{j}}$ from meson masses, including an extra binding of $-40.6\,$MeV for the $s\bar{s}$ system. These bindings are parametrized as functions of the reduced mass $\mu_{ij}$ with logarithmic scaling, generating a concave form $B(1/\mu)$ that satisfies the second-derivative condition (\ref{equ:soptheorem}). This fitting reveals a critical confinement scale of 1.34$\,$fm, beyond which bindings become positive, aligning with predictions for the string-breaking distance and indicating the onset of strong unquenched effects.

For baryonic systems, we adopt scaling ratios of $B_{ij} = B_{i\bar{j}}/2$ from color factors and $C_{ij} = 0.85 C_{i\bar{j}}$ to minimize the standard error to $\sigma = 2.07$~MeV across $\Delta M_{\textrm{EXP}} \approx \Delta B + \Delta C$. This close agreement confirms that the observed mass patterns are consistent with the concavity of binding energies and chromomagnetic interactions. Using the decomposition, we obtain approximate mass estimates for unobserved doubly and triply heavy baryons, as well as potential scattering channels like $D\bar{D} \to \eta_c + \pi$ and $\Xi_c^* \Xi_c^* \to \Xi^* + \Xi_{cc}^*$. Relations with large differences support physical pictures of the quark-exchange model, due to strong bindings and positive differences. These estimates are expected to be validated in future experiments.

\bigskip
\textbf{ACKNOWLEDGMENTS}

D. J. is supported by the National Natural Science Foundation of China under Grant No. 12165017.

\bigskip
\textbf{DATA AVAILABILITY}

No new experimental data were created in this study. The numerical codes and scripts used to generate the results are available at \href{https://github.com/SerialCore/MassInequality}{https://github.com/SerialCore/MassInequality}.


\begin{thebibliography}{99}
\bibitem{LHCb:2026pxn}
R. Aaij et al (LHCb Collaboration), arXiv:2603.28456 (2026)
\bibitem{Zhou:2025fpp}
H. Zhou, S. Q. Luo, X. Liu, Phys. Rev. D, 112 (7): 074007 (2025)
\bibitem{Bertlmann:1979zs}
R. A. Bertlmann, A. Martin, Nucl. Phys. B, 168: 111 (1980)
\bibitem{Richard:1983mu}
J. M. Richard, P. Taxil, Annals Phys., 150: 267 (1983)
\bibitem{Weingarten:1983uj}
D. Weingarten, Phys. Rev. Lett., 51: 1830 (1983)
\bibitem{Witten:1983ut}
E. Witten, Phys. Rev. Lett., 51: 2351 (1983)
\bibitem{Nussinov:1983hb}
S. Nussinov, Phys. Rev. Lett., 52: 966 (1984)
\bibitem{Lieb:1985aw}
E. H. Lieb, Phys. Rev. Lett., 54: 1987 (1985)
\bibitem{Martin:1986da}
A. Martin, J. M. Richard, P. Taxil, Phys. Lett. B, 176: 224 (1986)
\bibitem{Anwar:2017toa}
M. N. Anwar, J. Ferretti, F. K. Guo, et al., Eur. Phys. J. C, 78 (8): 647 (2018)
\bibitem{Karliner:2019vhw}
M. Karliner, J. L. Rosner, Phys. Rev. D, 101 (3): 036015 (2020)
\bibitem{Karliner:2014gca}
M. Karliner, J. L. Rosner, Phys. Rev. D, 90 (9): 094007 (2014)
\bibitem{Karliner:2017elp}
M. Karliner, J. L. Rosner, Nature, 551: 89 (2017)
\bibitem{Karliner:2017qjm}
M. Karliner, J. L. Rosner, Phys. Rev. Lett., 119 (20): 202001 (2017)
\bibitem{Ebert:2004ck}
D. Ebert, R. N. Faustov, V. O. Galkin, et al., Phys. Rev. D, 70: 014018 (2004), [Erratum: Phys. Rev. D 77, 079903 (2008)]
\bibitem{Yang:2019lsg}
G. Yang, J. Ping, P. G. Ortega, et al., Chin. Phys. C, 44 (2): 023102 (2020)
\bibitem{Liu:2019zoy}
Y. R. Liu, H. X. Chen, W. Chen, et al., Prog. Part. Nucl. Phys., 107: 237 (2019)
\bibitem{wu:2016vtq}
J. Wu, Y. R. Liu, K. Chen, et al., Phys. Rev. D, 97 (9): 094015 (2018)
\bibitem{luo:2017eub}
S. Q. Luo, K. Chen, X. Liu, et al., Eur. Phys. J. C, 77 (10): 709 (2017)
\bibitem{chen:2022asf}
H. X. Chen, W. Chen, X. Liu, et al., Rept. Prog. Phys., 86 (2): 026201 (2023)
\bibitem{Cheng:2020wxa}
J. B. Cheng, S. Y. Li, Y. R. Liu, et al., Chin. Phys. C, 45 (4): 043102 (2021)
\bibitem{li:2023aui}
S. Y. Li, Y. R. Liu, Z. L. Man, et al., Phys. Rev. D, 108 (5): 056015 (2023)
\bibitem{Zhang:2025wmr}
K. K. Zhang, W. X. Zhang, D. Jia, Phys. Rev. D, 112 (5): 054008 (2025)
\bibitem{Jia:2019bkr}
D. Jia, W. N. Liu, A. Hosaka, Phys. Rev. D, 101 (3): 034016 (2020)
\bibitem{Karliner:2020vsi}
M. Karliner, J. L. Rosner, Phys. Rev. D, 102 (9): 094016 (2020)
\bibitem{Song:2022csw}
Y. Song, D. Jia, W. Zhang, et al., Eur. Phys. J. C, 83 (1): 1 (2023)
\bibitem{Liu:2023vrk}
X. Y. Liu, W. X. Zhang, D. Jia, Phys. Rev. D, 108 (5): 054019 (2023)
\bibitem{Davies:2009tsa}
C. T. H. Davies, E. Follana, I. D. Kendall, et al (HPQCD Collaboration), Phys. Rev. D, 81: 034506 (2010)
\bibitem{Blossier:2010cr}
B. Blossier, P. Dimopoulos, R. Frezzotti, et al (ETM Collaboration), Phys. Rev. D, 82: 114513 (2010)
\bibitem{HPQCD:2011qwj}
R. J. Dowdall et al (HPQCD Collaboration), Phys. Rev. D, 85: 054509 (2012)
\bibitem{Davies:2012xun}
C. T. H. Davies, C. McNeile, A. Bazavov, et al., PoS, ConfinementX: 042 (2012)
\bibitem{Ebert:2002pp}
D. Ebert, R. N. Faustov, V. O. Galkin, Phys. Rev. D, 67: 014027 (2003)
\bibitem{Gregory:2009hq}
E. B. Gregory, C. T. H. Davies, E. Follana, et al., Phys. Rev. Lett., 104: 022001 (2010)
\bibitem{Mathur:2016hsm}
N. Mathur, M. Padmanath, R. Lewis, PoS, LATTICE2016: 100 (2016)
\bibitem{Mathur:2018epb}
N. Mathur, M. Padmanath, S. Mondal, Phys. Rev. Lett., 121 (20): 202002 (2018)
\bibitem{Bali:1998de}
G. S. Bali, arXiv:hep-ph/9809351 (1998)
\bibitem{Bali:2005fu}
G. S. Bali, H. Neff, T. Duessel, et al (SESAM Collaboration), Phys. Rev. D, 71: 114513 (2005)
\bibitem{Castorina:2007eb}
P. Castorina, D. Kharzeev, H. Satz, Eur. Phys. J. C, 52: 187 (2007)
\bibitem{Bulava:2019iut}
J. Bulava, B. H\"orz, F. Knechtli, et al., Phys. Lett. B, 793: 493 (2019)
\bibitem{Bonati:2020orj}
C. Bonati, S. Morlacchi, Phys. Rev. D, 101 (9): 094506 (2020)
\bibitem{Chagdaa:2021hul}
S. Chagdaa, B. Purev, E. Galsandorj, J. Phys. G, 48 (12): 125001 (2021)
\bibitem{Jiang:2023lmj}
J. J. Jiang, X. Chen, J. Qin, et al., Phys. Rev. D, 108 (12): 126002 (2023)
\bibitem{Kou:2024dml}
W. Kou, X. Chen, Phys. Lett. B, 856: 138942 (2024)
\bibitem{Lipkin:1986dx}
H. J. Lipkin, Phys. Lett. B, 171: 293 (1986)
\bibitem{Keren-Zur:2007ytk}
B. Keren-Zur, Annals Phys., 323: 631 (2008)
\bibitem{Brown:2014ena}
Z. S. Brown, W. Detmold, S. Meinel, et al., Phys. Rev. D, 90 (9): 094507 (2014)
\bibitem{Ebert:2005xj}
D. Ebert, R. N. Faustov, V. O. Galkin, Phys. Rev. D, 72: 034026 (2005)
\bibitem{LHCb:2017iph}
R. Aaij et al (LHCb Collaboration), Phys. Rev. Lett., 119 (11): 112001 (2017)
\bibitem{LHCb:2018pcs}
R. Aaij et al (LHCb Collaboration), Phys. Rev. Lett., 121 (16): 162002 (2018)
\bibitem{Barnes:2000hu}
T. Barnes, N. Black, E. S. Swanson, Phys. Rev. C, 63: 025204 (2001)
\bibitem{Hilbert:2007hc}
J. P. Hilbert, N. Black, T. Barnes, et al., Phys. Rev. C, 75: 064907 (2007)
\bibitem{Yamaguchi:2019djj}
Y. Yamaguchi, Y. Abe, K. Fukukawa, et al., EPJ Web Conf., 204: 01007 (2019)
\end{thebibliography}
\end{document}